\documentclass[utf8]{frontiersSCNSpreprint} 

\usepackage{url,hyperref,lineno,microtype,subcaption}
\usepackage[onehalfspacing]{setspace}
\usepackage{soul}
\usepackage{wrapfig}

\usepackage[textsize=scriptsize,backgroundcolor=green,disable]{todonotes}

\def\keyFont{\fontsize{8}{11}\helveticabold }
\def\firstAuthorLast{Pradhan {et~al.}} 
\def\Authors{Vivek Krishna Pradhan\,$^{1}$, Mike Schaekermann\,$^{2}$, Matthew Lease\,$^{1,*}$}
\def\Address{$^{1}$ University of Texas at Austin\\
$^{2}$ Amazon AI (work done prior to Amazon)}
\def\corrAuthor{Matthew Lease, School of Information, University of Texas at Austin}

\def\corrEmail{ml@utexas.edu}

\begin{document}
\onecolumn

\title[In Search of Ambiguity]{In Search of Ambiguity: A Three-Stage Workflow Design to Clarify Annotation Guidelines for Crowd Workers}

\author[\firstAuthorLast ]{\Authors} 
\address{} 
\correspondance{} 

\extraAuth{}

\maketitle

\begin{abstract}

We propose a novel three-stage FIND-RESOLVE-LABEL workflow for crowdsourced annotation to reduce ambiguity in task instructions and thus improve annotation quality. Stage 1 (FIND) asks the crowd to find examples whose correct label seems ambiguous given task instructions. Workers are also asked to provide a short tag which describes the ambiguous concept embodied by the specific instance found. We compare collaborative vs. non-collaborative designs for this stage. In Stage 2 (RESOLVE), the requester selects one or more of these ambiguous examples to label (resolving ambiguity). The new label(s) are automatically injected back into task instructions in order to improve clarity. Finally, in Stage 3 (LABEL), workers perform the actual annotation using the revised guidelines with clarifying examples. We compare three designs for using these examples: examples only, tags only, or both. We report image labeling experiments over six task designs using Amazon's Mechanical Turk. Results show improved annotation accuracy and further insights regarding effective design for crowdsourced annotation tasks. 

\tiny
 \keyFont{ \section{Keywords:} crowdsourcing, annotation, labeling, instructions, design, human-computer interaction, Mechanical Turk} 
\end{abstract}

\section{Introduction}

While crowdsourcing now enables labeled data to be obtained more quickly, cheaply, and easily than ever before 
\citep{snow2008cheap,alonso2015practical,sorokin2008utility}, ensuring data quality remains something of an art, challenge, and perpetual risk. Consider a typical workflow for annotating data on Amazon Mechanical Turk (MTurk): a {\em requester} designs an annotation task, asks multiple workers to complete it, and then post-processes labels to induce final consensus labels. Because the annotation work itself is largely opaque, with only submitted labels being observable, the requester typically has little insight into what if any problems workers encounter during annotation. While statistical aggregation  \citep{Sheshadri13, hung2013evaluation, zheng2017truth} 
and multi-pass iterative refinement \citep{little2010exploring, goto2016understanding} methods can be employed to further improve initial labels, there are limits to what can be achieved by post-hoc refinement following label collection. If initial labels are poor because many workers were confused by incomplete, unclear, or ambiguous  task instructions, there is a significant risk of ``garbage in equals garbage out'' \citep{vidgen2020directions}.

In contrast, consider a more traditional annotation workflow involving trusted annotators, such as practiced by the Linguistic Data Consortium (LDC) \citep{griffitt2016query}. Once preliminary annotation guidelines are developed, an iterative process ensues in which: 1) a subset of data is labeled based on current guidelines; 2) annotators review corner cases and disagreements, review relevant guidelines, and reach consensus on appropriate resolutions; 3) annotation guidelines are updated; and 4) the process repeats. In comparison to the simple crowdsourcing workflow above, this traditional workflow iteratively debugs and refines task guidelines for clarity and completeness in order to deliver higher quality annotations. 
However, it comes at the cost of more overhead, with a heavier process involving open-ended interactions with trusted annotators.
Could we somehow combine these for the best of both worlds?

In this work, we propose a novel three-stage FIND-RESOLVE-LABEL design pattern for crowdsourced annotation which strikes a middle-ground between the efficient crowdsourcing workflow on one hand and the high quality LDC-style workflow on the other. Like prior work \citep{bragg2018sprout,manam2018wingit,gaikwad2017daemo}, we seek to design a light-weight process for engaging the workers themselves to help debug and clarify the annotation guidelines.
However, existing approaches typically intervene in a reactive manner \textit{after} the annotation process has started, or tend to be constrained to a specific dataset or refinement of textual instruction only.
By contrast, our approach is proactive and open-ended.
It leverages crowd workers' unconstrained creativity and intelligence to identify ambiguous examples through semantic search on the Internet and enriches task instructions with these concrete examples proactively upfront before the annotation process commences.
Overall, we envision a partnership between requester and workers in which each party has complementary strengths and responsibilities in the annotation process, and we seek to maximize the relative strengths of each party to ensure data quality while preserving efficiency.

\begin{figure}[h]
  \centering
  \includegraphics[width=0.80\textwidth]{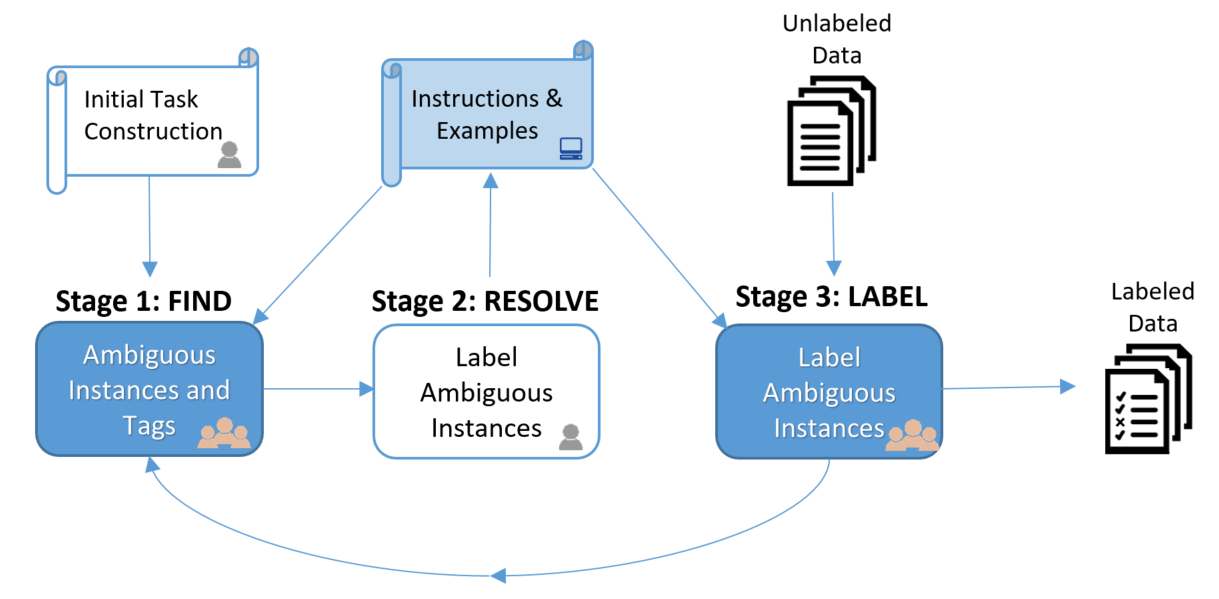}
    \caption{Our Three-Stage FIND-RESOLVE-LABEL workflow is shown above. Stage 1 (FIND) asks the crowd to find examples whose correct label seems ambiguous given the task instructions. In Stage 2 (RESOLVE), the requester selects and labels one or more of these ambiguous examples. These are then automatically injected back into task instructions in order to improve clarity. Finally, in Stage 3 (LABEL), workers perform the actual annotation using the revised guidelines with clarifying examples. If Stage 3 labeling quality is insufficient, we can return to Stage 1 to find more ambiguous examples to further clarify instructions. }
     \label{fig:flow}
\end{figure}

{\bf Figure \ref{fig:flow}} depicts our overall workflow. In Stage 1 (FIND), workers are shown initial guidelines for an annotation task and asked to search for data instances which appear ambiguous given the guidelines.  For each instance workers find, they are also asked to provide a short tag which describes the concept embodied by the specific instance which is ambiguous given the guidelines. Next, in Stage 2 (RESOLVE), the requester selects one or more of the ambiguous instances to label as exemplars. Those instances and their tags are then automatically injected back into the annotation guidelines in order to improve clarity. Finally, in Stage 3 (LABEL), workers perform the actual annotation using the revised guidelines with clarifying examples. The requester can run the LABEL stage on a sample of data, assess label quality, and then decide how to proceed. If quality is sufficient, remaining data can simply be labeled according to the guidelines. Otherwise, Stages 1 and 2 can be iterated in order to further refine the guidelines.

To evaluate our three-stage task design, we construct 6 different image labeling tasks with different levels of difficulty and intuitiveness. We construct a test dataset which contains different ambiguous and unambiguous concepts. Starting from simple and possibly ambiguous task instructions, we then improve instructions via our three-stage workflow. Given expert (gold) labels for our dataset for each of the 6 tasks, we can evaluate how well revised instructions compare to original instructions by measuring the accuracy of the labels obtained from the crowd.

{\bf Contributions}. We show that the crowd can find and provide useful ambiguous examples which can be used to further clarify task instructions, and that these examples can be utilized to improve annotation accuracy.
Our experiments further show that workers perform better when shown key ambiguous examples as opposed to randomly chosen examples.
Finally, we provide a sub-analysis on workers' performance for different intents of the same classification task and for different concepts of ambiguity within each intent.

Our paper is organized as follows.  \textbf{Section \ref{sec:motivation}} presents Motivation and Background. Next, \textbf{Section \ref{sec:workflow}} details our 3-Stage FIND-RESOLVE-LABEL workflow. Next, \textbf{Section \ref{sec:experimenalsetup}} explains our experimental setup. \textbf{Section \ref{sec:results}} then presents results. Finally, \textbf{Section \ref{sec:conandfuture}} discusses conclusions and future directions. 

\section{Motivation and Background}
\label{sec:motivation}
Consider the task of labeling images for object detection.
For example, on MTurk one might post a task such as, ``Is there a dog in this image?''
Such a task appears to be quite simple, but is it? For example, is a wolf a dog?
What about more exotic and unusual wild breeds of dogs?
Does the dog need to be a real animal or merely a depiction of one?
What about a museum model of an ancient but extinct dog breed, or a realistic wax sculpture
What if the dog is only partially visible in the image?
Ultimately, what is it that the requester really wants?
For example, a requester interested in anything and everything dog-related might have very liberal inclusion criteria.
On the other hand, a requester training a self-driving car might only care about animals to be avoided, while someone training a product search engine for an e-commerce site might want to include dog-style children's toys \citep{kulesza2014structured}.

As this seemingly simple example illustrates, annotation tasks that seem straightforward to a requester may in practice embody a variety of subtle nuances and ambiguities to be resolved.
Such ambiguities can arise for many reasons.
The requester may have been overly terse or rushed in posting a task.
They may believe the task is obvious and that no further explanation should be needed.
They likely also have their own implicit biases (of which they may be unaware) that provide a different internal conception of the task than others might have.
For example, the requester might be ignorant themselves of the domain (e.g., is a wolf a type of dog?) or have not fully-defined what they are looking for.
For example, in information retrieval, users' own conception and understanding of what they are looking for often evolves during the process of search and browsing \citep{cole2011theory}.
We describe our own experiences with this in \textbf{Section \ref{section:qualitative}}.
Annotators, on the other hand, also bring with them their own variety of implicit biases which the requester may not detect or understand \citep{ipeirotis2010quality, sen2015turkers, dumitrache2018crowdtruth, geva2019we, al-kuwatly-etal-2020-identifying, fazelpour2021diversity}. 

\subsection{Helping requesters Succeed}

{\bf Best practices.}
A variety of tutorials, surveys, introductions, and research papers offer how-to advice for successful microtask crowdsourcing with platforms such as MTurk \citep{kovashka2016crowdsourcing,egelman2014crowdsourcing,jones2013introduction,marshall2013experiences}.
For example, it is often recommended that requesters invest time browsing and labeling some data themselves before launching a task in order to better define and debug it \citep{alonso2015practical}. 
Studies have compared alternative task designs to suggest best practices  \citep{confusing,papoutsaki2015crowdsourcing,grady2010crowdsourcing,kazai2011crowdsourcing}. 

{\bf Templates and Assisted Design.}
Rather than start task design from scratch, MTurk offers templates and has suggested that requesters share successful templates for others' use \citep{chen2011opportunities}.
Similarly, classic research on software {\em design patterns} \citep{Gamma:1995:DPE:186897} has inspired ideas for similar crowdsourcing design patterns which could be reused across different data collection tasks.
For example, FIND-FIX-VERIFY \citep{bernstein2010soylent} is a well-known example that partially inspired our work.
Other researchers  have suggested improved tool support for workflow design \citep{kittur2012crowdweaver} or engaging the crowd itself in task design or decomposition \citep{kittur2011crowdforge,kulkarni2012collaboratively}.   

{\bf Automating Task Design.}
Other researchers have gone further still to propose new middleware and programmable APIs to let requesters define tasks more abstractly and leave some design and management tasks to the middleware \citep{barowy2016automan, chen2016mathematical, little2010turkit, ahmad2011jabberwocky, franklin2011crowddb}.

\subsection{Understanding Disagreement}

{\bf Random Noise vs.\ Bias.}
Since annotators are human, even trusted annotators will naturally make mistakes from time-to-time.
Fortunately, random error is exactly the kind of disagreement that aggregation \citep{Sheshadri13, hung2013evaluation, zheng2017truth} can easily resolve;
assuming such mistakes are relatively infrequent and independent, workers will rarely err at the same instance, and so techniques as simple as majority voting can address random noise.
On the other hand, if workers have individual biases, they will make consistent errors;
e.g., a teenager vs.\ a protective parent might have liberal vs.\ conservative biases in rating movies \citep{ipeirotis2010quality}.
In this case, it is useful to detect such consistent biases and re-calibrate worker responses to undo such bias.
Aggregation can also work provided that workers do not share the same biases.
However, when workers do share systematic biases, the independence assumption underlying aggregation is violated, and so aggregation can amplify bias rather than resolve it.
Consequently, it is important that task design annotation guidelines should be vetted to ensure they identify cases in which annotator biases conflict with desired labels and particularly establish clear expectations for how such cases should be handled \citep{draws2021bias, nouri2021iclarify}.

{\bf Objective vs.\ Subjective tasks.}
In fully-objective tasks, we assume each question has a single correct answer, and any disagreement with the gold standard reflects error.
Label aggregation methods largely operate in this space.
Tasks in which each question has a single correct answer.
At the other extreme, purely-subjective (i.e., opinion) tasks permit a wide range of valid responses with little expectation of agreement between individuals (e.g., asking one's favorite color or food).
Between these simple extremes, however, lies a wide, interesting, and important space of partially-subjective tasks in which answers are only partially-constrained \citep{tian2012learning,nguyen,sen2015turkers}.
For example, consider rating item quality: while agreement tends be high for items having extremely good or bad properties, instances with more middling properties naturally elicit a wider variance in opinion.
In general, because subjectivity permits a valid diversity of responses, it can be difficult to detect if an annotator does not undertake a task in good faith, complicating quality assurance.

{\bf Difficulty vs.\ Ambiguity.}
Some annotation tasks are more complex than others, just as some instances within each task are more difficult to label than other instances.
A common concern with crowdsourcing is whether inexpert workers have sufficient expertise to successfully undertake a given annotation task.
Intuitively, more guidance and scaffolding is likely necessary with more skilled tasks and less expert workers \citep{huang2021task}.
Alternatively, if we use sufficiently expert annotators, we assume difficult cases can be handled \citep{retelny2014expert,Vakharia15-iconf}.
With ambiguity, on the other hand, it would be unclear even to an expert what to do.
Ambiguity is an interaction between data instances and annotation guidelines;
effectively, an ambiguous instance is a corner-case wrt.\ guidelines. Aggregation can helpfully identify the majority interpretation, but that interpretation may or may not be what is actually desired.
Both difficult and ambiguous cases can lead to label confusion.
\cite{krivosheev2020confusedlabels} developed mechanisms to efficiently detect label confusion in classification tasks and demonstrated that alerting workers of the risk of confusion can improve annotation performance.

{\bf Static vs.\ Dynamic Disagreement.}
As annotators undertake a task, their understanding of work evolves as they develop familiarity with both the data and the guidelines.
In fact, prior work has shown that annotators interpret and implement task guidelines in different ways as annotation progresses \citep{scholer2013effect,kalra2017shifts}. 
Consequently, different sorts of disagreement can occur at different stages of annotation.
Temporally-aware aggregation can partially ameliorate this \citep{Jung15-hcomp}, as can implementing data collection processes to train, ``burn-in'', or calibrate annotators, controlling and/or accelerating their transition from an initial learning state into a steady state \citep{scholer2013effect}.
For example, we emphasize identifying key boundary-cases and expected labels for them.


\subsection{Mitigating Imperfect Instructions}

Unclear, confusing and ambiguous task instructions are commonplace phenomena on crowdsourcing platforms \citep{confusing,gadiraju2017clarity}.
In early work, \cite{alonso2008crowdsourcing}  recommended collecting optional, free-form, task-level feedback from workers.
While \citeauthor{alonso2008crowdsourcing} found that some workers did provide example-specific feedback, the free-form nature of their feedback request elicited a variety of response types, difficult to check or to invalidate spurious responses.
\citeauthor{alonso2008crowdsourcing} also found that requiring such feedback led many workers to submit unhelpful text that was difficult to automatically cull.
Such feedback was therefore made optional.

\cite{drapeau2016microtalk} proposed an asynchronous two-stage {\em Justify-Reconsider} method.
In the Justify task, workers provide a rationale along with their answer referring to the task guidelines taught during training.
For the Reconsider task, workers confronted with an argument for the opposing answer submitted by another worker and then asked to reconsider (i.e., confirm or change) their original answer.
The authors report that their Justify-Reconsider method generally yields higher accuracy, but that requesting justifications requires additional cost.
Consequently, they find that simply collecting more crowd annotations yields higher accuracy in a fixed-budget setting.

\cite{chang2017revolt} proposed a three-step approach in which crowd workers label the data, provide justifications for cases in which they disagree with others, and then review others' explanations.
They evaluate their method on an image labeling task and report that requesting only justifications (without any further processing) does not increase the crowd accuracy.
Their open-ended text responses can be subjective and difficult to check.

\cite{kulkarni2012mobileworks} provide workers a chat feature that supports workers in dealing with inadequate task explanations, suggesting additional examples to be given to requesters, teaching other workers how to use the UI, and verify their hypotheses of the underlying task intent.
\cite{schaekermann2018resolvable} investigate the impact of discussion among crowd workers on the label quality using a chat platform allowing synchronous group discussion.
While the chat platform allows workers to better express their justification than text-excerpts, the discussion increases task completion times.
In addition, chatting does not impose any restriction on topic, limiting discussion from unenthusiastic workers and efficacy.
\cite{chen2019cicero} also proposed a workflow allowing simultaneous discussion among crowd workers, and designed task instructions and a training phase to achieve effective discussions.
While their method yields high labeling accuracy, the increased cost due to the discussion limits its task scope.
\cite{manam2018wingit} evaluated both asynchronous and synchronous Q\&A between workers and requesters to allow workers to ask questions to resolve any uncertainty about overall task instructions or specific examples.
\cite{bragg2018sprout} proposed an iterative workflow in which data instances with low inter-rater agreement are put aside and either used as difficult training examples (if considered resolvable wrt. the current annotation guidelines) or used to refine the current annotation guidelines (if considered ambiguous).

Other work has explored approaches to address ambiguities even \textit{before} the annotation process commences.
For example, \cite{k2019taskmate} proposed a multi-step workflow enlisting the help of crowd workers to identify and resolve ambiguities in textual instructions.
\cite{gadiraju2017clarity} and \cite{nouri2021unclear} both developed predictive models to automatically score textual instructions for their overall level of clarity and \cite{nouri2021iclarify} proposed an interactive prototype to surface the predicted clarity scores to requesters in real-time as they draft and iterate on the instructions.
Our approach also aims to resolve ambiguities upfront, but focuses on identifying concrete visual examples of ambiguity and to automatically enrich the underlying set of textual instructions with those examples.

Ambiguity arises from the interaction between annotation guidelines and particular data instances.
Searching for ambiguous data instances within large-scale datasets or even the Internet can amount to finding a needle in a haystack.
There exists an analogous problem of identifying ``unknown unknowns'' or ``blind spots'' of machine learning models.
Prior has work has proposed crowdsourced or hybrid human-machine approaches for spotting and mitigating model blind spots \citep{Attenberg,vandenhof2019hybrid,lie2020uudetection}.
Our work draws inspiration from these workflows.
We leverage the scale, intelligence and common sense of the crowd to identify potential ambiguities within annotation guidelines and may thus aid in the process of mitigating blind spots in downstream model development.

\subsection{Crowdsourcing Beyond Data Labeling}

While data labeling represents the most common use of crowdsourcing in regard to training and evaluating machine learning models, human intelligence can be tapped in a much wider and more creative variety of ways.
For example, the crowd might verify output from machine learning models, identify and categorize blind spots \citep{Attenberg,vandenhof2019hybrid} and other failure modes \citep{cabrera2021discover}, and suggest useful features for a machine learning classifier \citep{cheng2015flock}.

One of the oldest crowdsourcing design patterns is utilizing the scale of the crowd for efficient, distributed exploration or filtering of large search spaces.
Classic examples include the search for extraterrestrial intelligence\footnote{\url{https://setiathome.berkeley.edu/}}, for Jim Gray's sailboat \citep{jimgray} 
or other missing people \citep{wang2010study}, for DARPA's red balloons \citep{pickard2011time}, for astronomical events of interest \citep{lintott2008galaxy}, for endangered wildlife \citep{rosser2019crowds} or bird species \citep{kelling2013human}, etc.
Across such examples, what is being sought must be broadly recognizable so that the crowd can accomplish the search task without need for subject matter expertise \citep{Kinney:2008:EDE:1458082.1458160}.
In the 3-stage FIND-FIX-VERIFY crowdsourcing workflow \citep{bernstein2010soylent}, the initial FIND stage  directs the crowd to identify ``patches'' in an initial text draft where more work is needed.

Our asking the crowd to search for ambiguous examples given task guidelines further explores the potential of this same crowd design pattern for distributed search.
Rather than waiting for ambiguous examples to be encountered by chance during the annotation process, we instead seek to rapidly identify corner-cases by explicitly searching for them.
We offload to the crowd the task of searching for ambiguous cases, and who better to identify potentially ambiguous examples than the same workforce that will be asked to perform the actual annotation?
At the same time we reduce requester work, limiting their effort to labeling corner-cases rather than adjusting the textual guidelines.

\section{Workflow Design}
\label{sec:workflow}
In this work, we propose a three-stage FIND-RESOLVE-LABEL workflow for clarifying ambiguous corner cases in task instructions.
An illustration of the workflow is shown in \textbf{Figure \ref{fig:flow}}.
In Stage 1 (FIND), workers are asked to proactively collect ambiguous examples and concept tags given task instructions (\textbf{Section \ref{sec:collectamb}}). 
Next, in Stage 2 (RESOLVE), the requester selects and labels one or more of the ambiguous examples found by the crowd.
These labeled examples are then automatically injected back into task instructions in order to improve clarity (\textbf{Section \ref{sec:rfeedback}}).
Finally, in Stage 3 (LABEL), workers perform the actual annotation task using the revised guidelines with clarifying examples (\textbf{Section \ref{sec:usingamb}}).
requesters run the final LABEL stage on a sample of data, assess label quality, and then decide how to proceed.
If quality is sufficient the remaining data can be labeled according to the current revision of the guidelines.
Otherwise, Stages 1 and 2 can be repeated in order to further refine the clarity of annotation guidelines.

\subsection{Stage 1: Finding Ambiguous Examples}
\label{sec:collectamb}

In Stage 1 (FIND), workers are asked to collect ambiguous examples given the task instructions.
For each ambiguous example, workers are also asked to generate a concept tag.
The concept tag serves multiple purposes.
First, it acts as a rationale \citep{Kutlu20-jair,mcdonnell2016relevant}, requiring workers to justify their answers and thus nudging them towards high-quality selections.
Rationales also provide a form of transparency to help requesters better understand worker intent.
Secondly, the concept tag provides a conceptual explanation of the ambiguity which can then be re-injected into annotation guidelines to help explain corner cases to future workers.

\begin{figure}[h]
  \centering
  \includegraphics[width=0.48\textwidth]{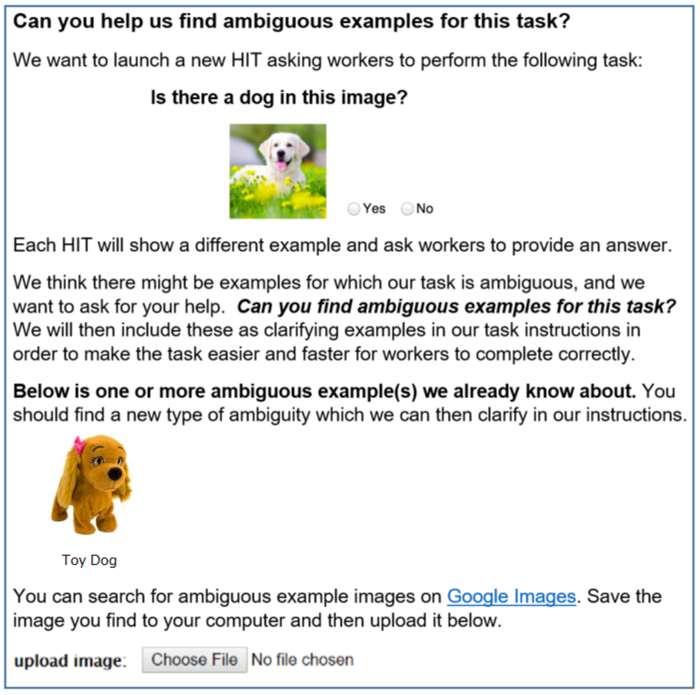}
    \caption{In the Stage 1 (FIND) task, workers are asked to search for examples they think would be ambiguous given task instructions. In this case, ``Is there a dog in this image?''  In collaboration conditions (Section \ref{section:collaboration}), workers will see additional ambiguous examples found by past workers.
    \\
    \\
    \\
    \\
    \\}
     \label{fig:colamb}
\end{figure}

{\bf Figure \ref{fig:colamb}} shows the main task interface for Stage 1 (FIND).
The interface presents the annotation task (e.g., \textit{``Is there a dog in this image?''}) and asks workers: \textit{``Can you find ambiguous examples for this task?''}
Pilot experiments revealed that workers had difficulty understanding the task based on this textual prompt alone.
We therefore make the additional assumption that requesters provide a single ambiguous example to clarify the FIND task for workers.
For example, the FIND stage for a dog annotation task could show the image of a Toy Dog as an ambiguous seed example.
Workers are then directed to use Google Image Search to find these ambiguous examples.
Once an ambiguous image is uploaded, another page (not shown) asks workers to provide a short concept tag summarizing the type of ambiguity represented by the example (e.g. \textit{Toy Dog}).

\textbf{Exploring Collaboration}:
\label{section:collaboration}
While crowd work is traditionally completed independently to prevent collusion and enable statistical aggregation of uncorrelated work \citep{Sheshadri13, hung2013evaluation, zheng2017truth}, a variety of work has explored collaboration mechanisms by which workers might usefully help each other complete a task more effectively \citep{dow2012shepherding,kulkarni2012mobileworks,drapeau2016microtalk,chang2017revolt,manam2018wingit,schaekermann2018resolvable,k2019taskmate,chen2019cicero}.
To investigate the potential value of worker collaboration in finding higher quality ambiguities, we explore a light-weight, iterative design in which workers do not directly interact with each other, but are shown examples found by past workers (in addition to the seed example provided by the requester).
For example, worker 2 would see an example selected by worker 1 and worker 3 would see examples found by workers 1 and 2, etc.
Our study compares three different collaboration conditions described in Section \ref{section:evaluate-stage1} below.


\subsection{Stage 2: Resolving Ambiguous Examples}
\label{sec:rfeedback}

\begin{figure}[h]
  \centering
  \includegraphics[width=0.48\textwidth]{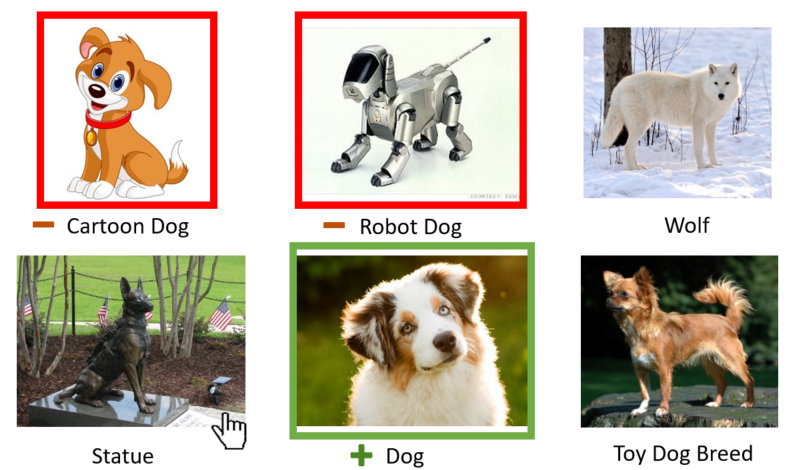}
  \caption{For Stage 2 (RESOLVE), our interface design lets a requester easily select and label images. Each mouse click on an example toggles between unselected, selected positive, and selected negative states.}
  \label{fig:reqUI}
\end{figure}

After collecting ambiguous examples in Stage 1 (FIND), the requester then selects and labels one or more of these examples.
The requester interface for Stage 2 (RESOLVE) is shown in {\bf Figure \ref{fig:reqUI}}.
Our interface design affords a low-effort interaction in which requesters toggle examples between three states via mouse click: (1) selected as positive example, (2) selected as negative example, (3) unselected.
Examples are unselected by default.
The selected (and labeled) examples are injected back into the task instructions for Stage 3 (LABEL).

\subsection{Stage 3: Labeling with Clarifying Examples}
\label{sec:usingamb}

\begin{figure}[h]
  \centering
  \includegraphics[width=0.48\textwidth]{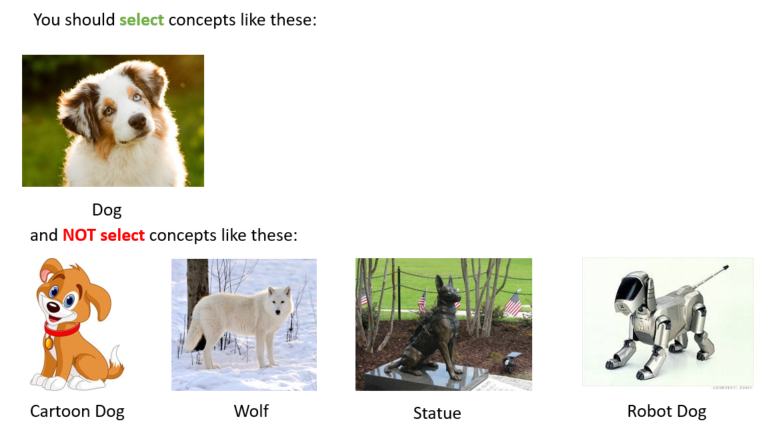}
  \caption{For Stage 3 (LABEL), we combine the ambiguous instances and/or tags collected in Stage 1 (FIND) with the requester labels from Stage 2 (RESOLVE) and automatically inject the labeled examples back into task instructions.}
  \label{fig:disamb}
\end{figure}

Best practices suggest that along with task instructions, requesters should include a set of examples and their correct annotations \citep{confusing}.
We automatically append to task instructions the ambiguous examples selected by the requester in Stage 2 (RESOLVE), along with their clarifying labels ({\bf Figure \ref{fig:disamb}}).
Positive examples are shown first (\textit{''you should select concepts like these''}), followed by negative examples (\textit{``and NOT select concepts like these''}).
Note that this stage simply concatenates the clarifying examples in this order and does not rely on task-specific instructions or layouts.

\section{Methods}
\label{sec:experimenalsetup}

Experiments were conducted in the context of image classification tasks.
In particular, we designed six annotation tasks representing different variations of labeling for the presence or absence of \textit{dog}-related concepts.
Similar to prior work \citet{kulesza2014structured}, we found this seemingly simplistic domain effective for our study because non-expert workers bring prior intuition as to how the classification could be done, but the task is characterized by a variety of subtle nuances and inherent ambiguities.
We employed a between-subjects design to avoid potential learning effects.
This design was enforced using ``negative'' qualifications \citep{negative-qualification} preventing crowd workers from participating in more than a single task.
For the purpose of experimentation, authors acted as requesters.
This included the specification of task instructions and intents and performing Stage 2 (RESOLVE), i.e., selecting clarifying examples for use in Stage 3 (LABEL).

\subsection{Dataset}
All experiments utilized the same set of 40 images.
The image set was designed to encompass both easy, unambiguous cases and a range of difficult, ambiguous cases with respect to the question \textit{``Is there a dog in this image?''}
We first assembled a set of candidate images using a combination of (1) online semantic image search conducted by the authors to identify a set of clear positive and clear negative examples, (2) the Stage 1 (FIND) mechanism in which crowd workers on Amazon's Mechanical Turk identified difficult, ambiguous cases.
Similar to \citet{kulesza2014structured}, we identified a set of underlying, dog-related categories via multiple passes of structured labeling on the data.
From this process, 11 categories of dog-related concepts emerged:
(1) dogs, (2) small dog breeds, (3) similar animals (easy to confuse with dogs), (4) cartoons, (5) stuffed toys, (6) robots, (7) statues, (8) dog-related objects (e.g., dog-shaped cloud), (9) miscellaneous (e.g., hot dog, the word ``dog''), (10) different animals (difficult to confuse with dogs) and (11) planes (the easiest category workers should never confuse with dogs).
Each image was assigned to exactly one category.

\subsection{Annotation Tasks}
\label{sec:intents}

When users of a search engine type in the query ``apple'', are they looking for information about the fruit, the company, or something else entirely? Despite the paucity of detail provided by a typical terse query, search result accuracy is assessed based on how well results match the user's  underlying intent. 
Similarly, requesters on crowdsourcing platforms expect workers to understand the annotation ``intent'' underlying the explicit instructions provided.
Analogously, worker accuracy is typically evaluated with respect to how well annotations match that requester intent even if instructions are incomplete, unclear or ambiguous.

To represent this common scenario, we designed three different annotation tasks.
For each task, the textual instructions exhibit a certain degree of ambiguity such that adding clarifying examples to instructions can help clarify requester intent to workers. 

For each of the three tasks, we also selected two different intents, one more intuitive than the other in order to assess the effectiveness of our workflow design under intents of varying intuitiveness.
In other words, we intentionally included one slightly more esoteric intent for each task hypothesizing that these would require workers to adapt to classification rules in conflict with their initial assumptions about requester intent.
For each intent below, we list the categories constituting the positive class.
All other categories are part of the negative class for the given intent intent.

For each of our six binary annotation tasks below, we partitioned examples into positive vs. negative classes given the categories included in the intent.
We then measured worker accuracy in correctly labeling images according to positive and negative categories for each task.

\subsubsection{Task 1: Is there a dog in this image?} 

\indent \textbf{Intent $a$} (more intuitive): dogs, small dog breeds 

\textbf{Intent $b$} (less intuitive): dogs, small dog breeds, similar animals. {\em Scenario}: The requester intends to train a machine learning model for avoiding animals and believes the model may also benefit from detecting images of wolves and foxes. 

\subsubsection{Task 2: Is there a fake dog in this image?} 

\indent \textbf{Intent} $a$ (more intuitive): similar animals.
{\em Scenario}: The requester is looking for animals often confused with dogs.

\textbf{Intent} $b$ (less intuitive): cartoons, stuffed toys, robots, statues, objects. 
{\em Scenario}: The requester is looking for inanimate objects representing dogs.

\subsubsection{Task 3: Is there a toy dog in this image?}

\indent \textbf{Intent $a$} (less intuitive): small dog breeds. {\em Scenario}: Small dogs, such as Chihuahua or Yorkshire Terrier, are collectively referred to as ``toy dog'' breeds\footnote{\url{https://en.wikipedia.org/wiki/Toy_dog}}. 
However, this terminology is not necessarily common knowledge making this intent less intuitive.

\textbf{Intent $b$} (more intuitive): stuffed toys, robots. {\em Scenario}: The requester is looking for children's toys, e.g., to train a model for an e-commerce site.

\subsection{Evaluation}

\subsubsection{Qualitative Evaluation of Ambiguous Examples from Stage 1 (FIND)}
\label{section:evaluate-stage1}

For Stage 1 (FIND), we evaluated crowd workers' ability to find ambiguous images and concept tags for Task 1: \textit{``Is there a dog in this image?''}.
Through qualitative coding, we analyzed worker submissions based on three criteria:
(1) correctness; (2) uniqueness; and (3) usefulness.

{\em Correctness} captures our assessment of whether the worker appeared to have understood the task correctly and submitted a plausible example of ambiguity for the given task.
Any incorrect examples were excluded from consideration for uniqueness or usefulness.

{\em Uniqueness} captures our assessment of how many distinct types of ambiguity workers found across correct examples.
For example, we deemed ``Stuffed Dog'' and ``Toy Dog'' sufficiently close as to represent the same concept.

{\em Usefulness} captures our assessment of which of the unique ambiguous concepts found were likely to be useful in annotation.
For example, while an image of a hot dog is valid and unique, it is unlikely that many annotators would find it ambiguous in practice.

Our study compares three different collaboration conditions for Stage 1 (FIND).
In all three conditions, workers were shown one or more ambiguous examples with associated concept tags and were asked to add another, different example of ambiguity, along with a concept tag for that new example:

\begin{enumerate}
    \item \textbf{No Collaboration}.
Each worker sees the task interface seeded with a single ambiguous example and its associated concept tag provided by the requester.
Workers find additional ambiguous examples \textit{independently} from other workers.

\item \textbf{Collaboration}.
Workers see all ambiguous examples and their concept tags previously found by other workers.
There is no filtering mechanism involved, so workers may be presented with incorrect and/or duplicated examples.
This workflow configuration amounts to a form of unidirectional, asynchronous communication among workers.
 
\end{enumerate}

For all three collaboration conditions a total of 15 ambiguous examples (from 15 unique workers) were collected and evaluated with respect to the above criteria.

\subsubsection{Quantitative Evaluation of Example Effectiveness in Stage 3 (LABEL)}

To evaluate the effectiveness of enriching textual instructions with ambiguous examples from Stage 1 (FIND) and to assess the relative utility of presenting workers with images and/or concept tags from ambiguous examples, we compared the following five conditions.
The conditions varied in how annotation instructions were presented to workers in Stage 3 (LABEL):
\begin{enumerate}
    \item \textbf{B0}: No examples were provided along with textual instructions.
    \item \textbf{B1}: A set of randomly chosen examples were provided along with textual instructions.
    \item \textbf{IMG}: Only images (but no concept tags) of ambiguous examples were shown to workers along with textual instructions.
    \item \textbf{TAG}: Only concept tags (but no images) of ambiguous examples were shown to workers along with textual instructions.
    \item \textbf{IMG+TAG}: Both images and concept tags of ambiguous examples were shown to workers along with textual instructions.
\end{enumerate}

Each of the five conditions above was completed by 9 unique workers.
Each task consisted of classifying 10 images.
Workers were asked to classify each of the 10 images into either the positive or the negative class.

\section{Results}
\label{sec:results}
\subsection{Can Workers Find Ambiguous Concepts?}

In this section, we provide insights from pilots of Stage 1 (FIND) followed by a qualitative analysis of ambiguous examples identified by workers in this stage.

\subsubsection{Pilot Insights}
\label{section:qualitative}

\textbf{Task Design:}
Initial pilots of Stage 1 (FIND) revealed two issues: 1) duplicate concepts, and 2) misunderstanding of the task.
Some easy-to-find and closely related concepts were naturally repeated multiple times.
One type of concept duplication was related to the \textit{seed} example provided by requesters to clarify the task objective.
In particular, some workers searched for additional examples of the \textit{same} ambiguity rather than finding \textit{distinct} instances of ambiguity.
Another misunderstanding led some workers to submit \textit{generally} ambiguous images, i.e., similar to Google Image Search results for search term ``ambiguous image'', rather than images that were ambiguous relative to the specific task instruction \textit{``Is there a dog in this image?''}
We acknowledge that our own task design was not immune to ambiguity, so we incorporated clarifications to instruct workers to find ambiguous examples \textit{distinct} from the seed example and \textit{specific} to the task instructions provided.

\begin{figure}[h]
  \centering
    \includegraphics[width=0.48\textwidth]{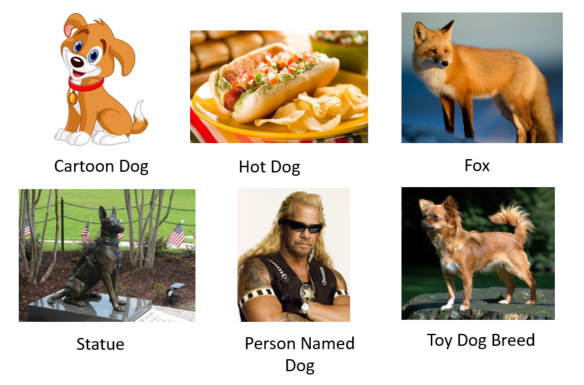}
    \caption{Ambiguous examples and concept tags provided by workers in Stage 1 (FIND) for the task ``Is there a dog in this image?''. We capitalize tags here for presentation but use raw worker tags without modification in our evaluation.}
    \label{fig:ambex}
\end{figure}

\textbf{Unexpected Ambiguous Concepts:}
However, our pilots also revealed workers' ability to identify surprising examples of ambiguous concepts unanticipated by the paper authors.
Some of these examples were educational and helped the paper authors learn about the nuances of our task.
For example, one worker returned an image of a Chihuahua (small dog breed) along with the concept tag ``toy dog.''
In trying to understand the worker's intent we learned that the term ``toy dog'' is a synonym of small dog breeds. \footnote{\url{https://en.wikipedia.org/wiki/Toy_dog}}
Prior to that, our interpretation of the ``toy dog'' concept was limited to children's toys.
This insight inspired Task 3 (``Is there a toy dog in this image?'') with two different interpretations (\textbf{Section \ref{sec:intents}}).
Another unexpected ambiguous example was the picture of a man ({\bf Figure \ref{fig:ambex}}).
We initially jumped to the conclusion that the worker's response was spam, but on closer inspection discovered that the picture displayed reality show celebrity ``Dog the Bounty Hunter.'' \footnote{\url{http://www.dogthebountyhunter.com}}
These instances are excellent illustrations of the possibility that crowd workers may interpret task instructions in valid and original ways entirely unanticipated by requesters.

\subsubsection{Qualitative Assessment of Example Characteristics}

We employed qualitative coding to assess whether worker submissions met each of the quality criteria (Correctness, Uniqueness, Usefulness).
\textbf{Table \ref{tab:collab}} shows the percentage of ambiguous examples meeting these criteria for the two conditions with and without collaboration respectively.
Our hypothesis that collaboration among workers can help produce higher quality ambiguous examples is supported by our results.
Results show that, compared to no collaboration, a collaborative workflow produced substantially greater proportions of correct (93\% vs. 60\%), unique (40\% vs. 27\%) and useful (33\% vs. 27\%) ambiguous examples.
A potential explanation for this result is that exposing workers to a variety of ambiguous concepts upfront may assist them in exploring the space of yet uncovered ambiguities more effectively.

\subsection{Can Ambiguous Examples Improve Annotation Accuracy?}

Next, we report quantitative results on how ambiguous examples--found in Stage 1 and selected and labeled in Stage 2--can be used as instructional material to improve annotation accuracy in Stage 3.
We also provide an analysis of annotation errors. 

\subsubsection{Effectiveness of Ambiguous Examples}

Our hypothesis is that these examples can be used to help delineate the boundary of our annotation task, and hence teach annotation guidelines to crowd workers better than randomly chosen examples.
{\bf Table \ref{tab:mainres}} reports crowd annotation accuracy for each of the six tasks broken down by experimental condition.

\textbf{Using Examples to Teach Annotation Guidelines:}
Intuitively, providing examples to workers helps them to better understand the intended labeling task \citep{confusing}.
Comparing designs B0 and B1 in \textbf{Table \ref{tab:mainres}}, we clearly see that providing examples (B1) almost always produces more accurate labeling than a design that provides no examples (B0).
In addition to this, the IMG design performs better than B1.
This shows that the kind of examples that are provided is also important. Showing ambiguous examples is clearly superior to showing randomly chosen examples.
This supports our hypothesis: ambiguous examples appear to delineate labeling boundaries for the task better than random examples.

\textbf{Instances vs.\ Concepts:}
Best practices suggest that requesters provide examples when designing their task \citep{confusing}.
We include this design in our evaluation as B1. An alternate design is to show concepts as examples instead of specific instances; this is our design TAG, shown in \textbf{Table \ref{tab:mainres}}. For example, for the task ``Is there a Dog in this image?'', instead of showing a dog statue image, we could simply provide the example concept ``Inanimate Objects'' should be labeled as NO. Results in \textbf{Table \ref{tab:mainres}} show that TAG consistently outperform IMG, showing that teaching via example concepts can be superior to teaching via example instances.

\textbf{Concepts only vs.\ Concepts and Examples:}
Surprisingly, workers shown only concept tags performed better than workers shown concept tags along with an example image for that concept. Hence the particular instance chosen may not represent the concept well. This might be overcome  by better selecting a more representing example for a concept, or showing more examples for each concept. We leave such questions for future work.

\subsubsection{Sources of Worker Errors}

\textbf{Difficult vs.\ Subjective Questions:}

{\bf Table \ref{tab:difftask}} shows accuracy for categories ``Similar Animal'' and ``Cartoon'' for Task 1b (\textbf{Section \ref{sec:intents}}). We see that some concepts appear more difficult, such as correctly labeling a wolf or a fox. Annotators appear to need some world knowledge or training of differences between species in order to correctly distinguish such examples vs.\ dogs. Such concepts seem more difficult to teach; even though the accuracy improves, the improvement is less than we see with other concepts. In contrast, for Cartoon Dog (an example of a subjective question), adding this category to the illustrative examples greatly reduces the ambiguity for annotators. Other concepts like ``Robot'' and ``Statue'' also show large improvements in accuracy.

\textbf{Learning Closely Related Concepts:}
To see if crowdworkers learn closely related concepts without being explicitly shown examples, consider ``Robot Dog'' and ``Stuffed Toy'' as two types of a larger ``Toy Dog'' children's toy concept. In Task 1b, the workers are shown the concept ``Robot Dog'' as examples labeled as NO, without being shown an example  for ``Stuffed Toy''. 
\textbf{Table \ref{tab:difftask}} shows that workers learn the related concept ``Stuffed Toy'' and accurately label the instances that belong to this concept. The performance gain for the concept ``Toy Dog'' is the same as the gain for ``Robot Dog'', when we compare design IMG+TAG and B1. Other similarly unseen concepts (marked with an asterisk(*) in the table) show that workers are able to learn the requester's intent for unseen concepts if given examples of other, similar concepts.

\textbf{Peer Agreement with Ambiguous Examples:}
It is not always possible or cost-effective to obtain expert/gold labels for  tasks, so requesters often rely on peer-agreement between workers to estimate worker reliability. Similarly, majority voting or weighted voting is often used to aggregate worker labels for consensus \citep{Sheshadri13, hung2013evaluation, zheng2017truth}. However, we also know that when workers have consistent, systematic group biases, aggregation will serve to reinforce and amplify the group bias rather than mitigate it \citep{ipeirotis2010quality,sen2015turkers,dumitrache2018crowdtruth, fazelpour2021diversity}.
While we find agreement often correlates with accuracy, and so have largely omitted reporting it in this work, we do find a number of concepts for which the majority chooses wrong answers, producing high agreement but low accuracy. Recall that our results are reported over 9 workers per example, whereas typical studies use a plurality of 3 or 5 workers. Also recall that Tasks 1b and 2a (\textbf{Section \ref{sec:intents}}) represent two of our less intuitive annotation tasks for which requester intent may be at odds with worker intuition, requiring greater task clarity for over-coming worker bias. 
{\bf Table \ref{tab:agree}} shows majority vote accuracy for these tasks for the baseline B1 design which (perhaps typical of many requesters)  includes illustrative examples but not necessarily the most informative ones.  Despite collecting labels from 9 different workers, the majority is still wrong, with majority vote accuracy on ambiguous examples falling below 50\%.

\section{Conclusion and Future Work}
\label{sec:conandfuture}

\subsection{Summary and Contributions} 

While crowdsourcing has rendered the process of data annotation straightforward, rapid, and low-cost, data quality assurance remains a challenge today. We posit that in adopting the many advantages crowdsourcing has to offer some researchers may be tempted to neglect quality assurance best practices from traditional export annotation workflows.
Our work adds to the existing literature on mechanisms to disambiguate nuanced class categories and thus improve labeling guidelines and, in effect, classification decisions in crowdsourced data annotation.

In this work, we presented a three-stage FIND-RESOLVE-LABEL workflow as a novel mapping of traditional annotation processes, involving iterative refinement of guidelines by expert annotators, onto a light-weight, structured task design suitable for crowdsourcing.
Through careful task design and intelligent distribution of effort between crowd workers and requesters, we have shown that the crowd can play a valuable role in reducing requester effort while also helping requesters to better understand the nuances and edge cases of their intended annotation taxonomy in order to generate clearer task instructions for the crowd.
In contrast to prior work, our approach is proactive and open-ended, leveraging crowd workers' unconstrained creativity and intelligence to identify ambiguous examples through open-ended semantic search on the Internet, proactively enriching task instructions with these examples upfront before the annotation process commences.

While including illustrative examples in instructions is known to be helpful \citep{confusing}, we have shown that not all examples are equally informative to annotators, and that intelligently selecting ambiguous corner-cases can improve labeling quality.
Our results revealed that the crowd performed worst on ambiguous instances and thus can benefit the most from help for cases where requester intents run counter to annotators' internal biases or intuitions.
For some instances of ambiguity, we observed high agreement among workers on answers contrary to what the requester defines as correct.
Such tasks are likely to produce an incorrect label even when we employ intelligent answer aggregation techniques. Techniques like ours to refine instruction clarity are particularly critical in such cases.

Finally, we found that workers were able to infer the correct labels for concepts closely related to the target concept.
This result suggests that it may not be necessary to identify and clarify \textit{all} ambiguous concepts that could potentially be encountered during the task. An intelligently selected set of clarifying examples may enable the crowd to disambiguate labels of unseen examples accurately even if not all instances of ambiguity are exhaustively covered.

\subsection{Future Work} 

While we evaluate our strategy on an image labeling task, our approach is more general and could be usefully extended to other domains and tasks.
For example, consider collecting document relevance judgments in information retrieval \citep{alonso2008crowdsourcing,scholer2013effect,mcdonnell2016relevant}, where user {\em information needs} are often subjective, vague, and incomplete.
Such generalization may raise new challenges.
For example, our the image classification task used in our study allows us to point workers to online semantic image search.
However, other domains may require additional or different search tools for workers to be able to effectively identify ambiguous corner cases.

\citet{alonso2015practical} proposes having workers perform practice tasks to get familiarized with the data and thus increase annotation performance for future tasks.
While our experimental setup prevented workers from performing more than one task to avoid potential learning effects, future work may explore and leverage workers' ability to improve their performance for certain types of ambiguity over time.
For example, we may expect that workers who completed Stage 1 are better prepared for Stage 3 given that they have already engaged in the mental exercise of critically exploring the decision boundary of for the class taxonomy in question.

Another best practice from LDC is deriving a decision tree 
for common ambiguous cases which annotators can follow as a principled and consistent way to determine the label of ambiguous examples \citep{griffitt2016query}.
How might we use the crowd to induce such a decision tree?  Prior design work in effectively engaging the crowd in clustering \citep{chang2016alloy} can guide design considerations for this challenge.

In our study, Stage 2 RESOLVE required requesters to select ambiguous examples.
Future work may explore variants of Stage 1 FIND where requesters \textit{filter} the ambiguous examples provided by the crowd.
There is an opportunity for saving requester effort if both of these stages are combined.
For instance, examples selected in the filtering step of Stage 1 can be fed forward to reduce the example set considered for labeling in Stage 2.
Another method would be to have requesters perform labeling simultaneously with filtering in Stage 1, eliminating Stage 2 altogether.
Finally, if the requester deems the label quality of Stage 3 insufficient and initiates another cycle of ambiguity reduction via Stages 1 and 2 those stages could start with examples already identified in the prior cycle.

A variety of other directions can be envisioned for further reducing requester effort.
For example, the crowd could be called upon to verify and prune ambiguous examples collected in the initial FIND stage.
Examples flagged as spam or assigned a low ambiguity rating could be automatically discarded to minimize requester involvement in Stage 2.
Crowd ambiguity ratings could also be used to rank examples for guiding requesters' attention in Stage 2.
A more ambitious direction for future work would be to systematically explore how well and under what circumstances the crowd is able to correctly infer requester intent.
Generalizable insights about this question would enable researchers to design strategies that eliminate requester involvement altogether under certain conditions.


\section*{Conflict of Interest Statement}

The authors declare that the research was conducted in the absence of any commercial or financial relationships that could be construed as a potential conflict of interest. While Matthew Lease is engaged as an Amazon Scholar, proposed methods are largely general across paid and volunteer labeling platforms.

\section*{Author Contributions}

VKP: implementation, experimentation, data analysis, manuscript preparation; MS: manuscript preparation; ML: advising on research design and implementation, manuscript preparation.

\section*{Funding}
This research was supported in part by the Micron Foundation and by Good Systems (\url{https://goodsystems.utexas.edu}), a UT Austin Grand Challenge to develop responsible AI technologies. Any opinions, findings, and conclusions or recommendations expressed by the authors are entirely their own and do not represent those of the sponsors.

\section*{Acknowledgments}
We thank our many talented crowd contributors, without whom our research would not be possible. We thank the reviewers for their time and assistance in helping us to further improve this work.


\bibliographystyle{frontiersinSCNS_ENG_HUMS} 
\bibliography{bibliography}


\newpage
\section*{Figures and Tables}


\begin{table}[h]
\centering
\caption{Percentage of correct, unique and useful examples from Stage 1 (FIND).}
\label{tab:collab}
\begin{tabular}{|l|c|c|c|}
\hline
                         & \textbf{Correct}       & \textbf{Unique}        & \textbf{Useful}          \\ \hline
No Collaboration         & 60.0          & 26.7        & 26.7          \\ \hline
Collaboration & \textbf{93.0} & \textbf{40.0} & \textbf{33.3} \\ \hline
\end{tabular}
\end{table}

\begin{table}[h]
\centering
\caption{Worker accuracy [\%] for all six tasks by condition.} 
\label{tab:mainres}
\begin{tabular}{|l|c|c|c|c|c|c|}
\hline
\textbf{Design} & \textbf{Task 1a} & \textbf{Task 1b} & \textbf{Task 2a} & \textbf{Task 2b} & \textbf{Task 3a} & \textbf{Task 3b} \\ \hline
B0              & 75.6             & 70.1             & 47.8             & 78.7             & 69.1             & 92.9             \\ \hline
B1              & 83.0             & 66.4             & 59.0             & 85.5             & 78.7             & 96.0             \\ \hline
IMG             & 88.0             & 89.5             & 68.5             & 85.8             & 89.2             & 93.5             \\ \hline
TAG             & 91.0             & \textbf{91.0}    & 79.0             & \textbf{87.0}    & \textbf{91.0}    & \textbf{96.9}    \\ \hline
IMG+TAG         & \textbf{91.4}    & 87.0             & \textbf{81.8}    & 86.4             & 88.3             & 96.6             \\ \hline
\end{tabular}

\end{table}

\begin{table}[h]
\centering
\caption{Worker accuracy [\%] on Task 1b, broken down by concept category. We see that some hard concepts cannot be easily disambiguated, e.g.\ Similar Animal. Concepts for which workers were not shown any examples are marked with an asterisk(*).}
\label{tab:difftask}
\begin{tabular}{|l|c|c|c|c|c|c|}
\hline
\textbf{Design} & \textbf{\begin{tabular}[c]{@{}c@{}}Similar\\ Animal\end{tabular}} & \textbf{\begin{tabular}[c]{@{}c@{}}Stuffed\\ Toy*\end{tabular}} & \textbf{Robot} & \textbf{Statue} & \textbf{Cartoon} & $\dotsb$ \\ \hline
B0                               & 37.0                                                                                                                                                                                                                              & 22.2                                                                                                                                                     & 33.3                            & 18.5                             & 62.2                                                                                                                                                       &                                \\ \hline
B1                               & 14.8                                                                                                                                                                                                                              & 22.2                                                                                                                                                     & 25.9                            & 29.6                             & 57.8                                                                                                                                                       &                                \\ \hline
IMG                              & 44.4                                                                                                                                                                                                                              & {\bf 100.0}                                                                                                                             & {\bf 100.0}    & 92.6                             & {\bf 100.0}                                                                                                                               &        \\ \hline
TAG                              & {\bf 74.1}                                                                                                                                                                                                       & {\bf 100.0}                                                                                                                             & 88.9                            & {\bf 88.9}      & 97.8                                                                                                                                                       &        \\ \hline
{\tiny IMG+TAG} & 48.1                                                                                                                                                                                                                              & 88.9                                                                                                                                                     & 92.6                            & 77.8                             & 97.8                                                                                                                                                       &                               \\ \hline

\textbf{Design} & \textbf{Objects*} & \textbf{\begin{tabular}[c]{@{}c@{}}Unseen\\ \!\!Ambiguity*\!\!\end{tabular}} & \textbf{\begin{tabular}[c]{@{}c@{}}Small\\  Breed\end{tabular}} & \textbf{Plane} & \textbf{Dog} & \textbf{\begin{tabular}[c]{@{}c@{}}Another\\ Animal\end{tabular}}  \\ \hline
B0                               & 74.1 & {\bf 88.9}                                                                                                                                                                                                       & 88.9                                                                                                                                                     & {\bf 100.0}    & {\bf 100.0}     & {\bf 100.0}                                                                                                                                                                   \\ \hline
B1                               & 59.3 & 82.2                                                                                                                                                                                                                              & {\bf 94.4}                                                                                                                              & {\bf 100.0}    & {\bf 100.0}     & {\bf 100.0}                                                                                                                                                                   \\ \hline
IMG                              & {\bf 100.0} & 75.6                                                                                                                                                                                                                              & 88.9                                                                                                                                                     & {\bf 100.0}    & 95.6                             & {\bf 100.0}                                                                                                                                                                   \\ \hline
TAG                              & {\bf 100.0} & 80.0                                                                                                                                                                                                                              & {\bf 94.4}                                                                                                                              & {\bf 100.0}    & 91.1                             & {\bf 100.0}                                                                                                                                                                   \\ \hline
{\tiny IMG+TAG} & 96.3 & 77.8                                                                                                                                                                                                                              & 91.7                                                                                                                                                     & 96.3                            & 95.6                             & 88.9                                                                                                                                                                                           \\ \hline
\end{tabular}
\end{table}

\begin{table}[h]
\centering

\caption{Worker accuracy [\%] on ambiguous vs unambiguous categories with baseline B1. For Tasks 1b and 2a, the majority vote (among 9 workers) on ambiguous examples is wrong.}
\label{tab:agree}
\begin{tabular}{|l|c|c|}
\hline
\textbf{Task} & \textbf{Unambiguous}            & \textbf{Ambiguous}            \\ \hline
1a   & 96.1                   & 72.2                 \\ \hline
1b   & 98.6                   & \textbf{41.7}                 \\ \hline
2a   & 86.8                   & \textbf{42.2}                 \\ \hline
2b   & 98.5                   & 82.5                 \\ \hline
3a   & 82.5                   & 75.8                 \\ \hline
3b   & 98.6                   & 93.8                 \\ \hline
\end{tabular}

\end{table}


\end{document}